# Composition-spread Growth and the Robust Topological Surface State of Kondo insulator $SmB_6$ Thin Films

Jie Yong[1,2,*], Yeping Jiang[1,2,*], Demet Usanmaz[3], Stefano Curtarolo[3], Xiaohang Zhang[1,2], Linze Li[4], Xiaoqing Pan[4], Jongmoon Shin[5], Ichiro Tachuchi[1,5], Richard L. Greene[1,2]

[1]Center for Nanophysics & Advanced Materials, University of Maryland, College Park, Maryland 20742, USA

[2]Department of physics, University of Maryland, College Park, Maryland 20742, USA

[3]Department of Mechanical Engineering and Materials Science, Duke University, Durham, NC 27708

[4]Department of Materials Science and Engineering, University of Michigan, Ann Arbor, Michigan 48109, USA

[5]Department of Materials Science & Engineering, University of Maryland, College Park, Maryland 20742, USA

Topological insulators are a class of materials with insulating bulk but protected conducting surfaces due to the combination of spin-orbit interactions and time-reversal symmetry. The surface states are topologically non-trivial and robust against non-magnetic backscattering, leading to interesting physics and potential quantum computing applications[1, 2]. Recently there has been a fast growing interest in samarium hexboride ($SmB_6$), a Kondo insulator predicted to be the first example of a correlated topological insulator[3, 4]. Here we fabricated smooth thin films of nanocrystalline $SmB_6$ films. Their transport behavior indeed shows that $SmB_6$ is a bulk insulator with topological surface states. Upon decreasing the temperature, the resistivity ρ of $Sm_{0.14}B_{0.86}$ ($SmB_6$) films display significant increase below 50 K due to hybridization gap formation, and it shows a saturation behavior below 10 K. The saturated resistance of our textured films is similar to that of the single crystals, suggesting that this conduction is from the surface and robust against grain boundary scatterings. Point contact spectroscopy (PCS) of the film using a superconducting tip displays both a Kondo Fano resonance and Andreev reflection, suggesting the existence of both an insulating Kondo lattice and metallic surface states.

[a] Electronic mail: jyong@umd.edu ypjiang@umd.edu

*These authors contributed equally to this work

Samarium hexboride has recently been predicted to be the first example of topological Kondo insulator[3, 4]. Upon decreasing the temperature, the resistivity of $SmB_6$ increases like an insulator but saturates at temperatures below 5K [ref. 5-14]. Recent transport measurements[8-15] show that the resistance of this saturation is thickness-independent for both longitudinal and transverse directions[8-10]. Moreover, weak anti-localization and linear magnetoresistance have been observed in single crystal $SmB_6$ to support the presence of spin-momentum locked surface states[12]. It has been shown that doping $SmB_6$ with magnetic impurities diminishes this saturation, while non-magnetic impurities do not[9]. These facts suggest that the conduction is from the surface state, it is protected by time-reversal symmetry, and is robust against non-magnetic scatterings. Quantum oscillations have been observed by torque magnetometry measurements[16] and the tracking of the Landau levels in the infinite magnetic field limit points to -1/2, which indicates a 2D Dirac electronic state. Surface sensitive measurements such as angle resolved photoemission (ARPES) [17-23], scanning tunneling microscopy[24-26] (STM) confirm the formation of hybridization gap and the existence of the surface states. One study[23] in particular suggests that the surface states are spin polarized with spin momentum locking which is a signature of the topological states. Preparation of $SmB_6$ in thin film geometries with smooth surfaces can facilitate device patterning and surface sensitive measurements that will give direct insight on the topological nature of this material. In particular, smooth thin films would allow fabrication of heterostructures for probing the long-sought Majorana fermions by coupling this material with a conventional superconductor[27].

In this paper, we report the growth of $SmB_6$ films by co-sputtering of $SmB_6$ and boron targets. Composition and structural analyses show that our films are polycrystalline $SmB_6$ films with smooth surfaces. Transport and point contact spectroscopy (PCS) measurements confirmed a robust surface conduction at low temperatures with an insulating bulk, suggesting a topological behavior as found in single crystal $SmB_6$.

We initially attempted growth of $SmB_6$ thin films via pulsed laser deposition and sputtering using single $SmB_6$ targets, but these growths resulted in grossly boron deficient films. In order to achieve the correct stoichiometry, we turned to the combinatorial composition-spread approach[28]. By covering a large range of continuous average composition on a given spread wafer, this method can be used to ensure that somewhere on the wafer, there is a region with the correct average stoichiometry (see Methods for growth conditions).

The composition variation of the $Sm_xB_{1-x}$ spreads is mapped by wavelength dispersive spectroscopy (WDS), with an error bar of about 1%. Fig. 1a plots the composition of Sm(x) mapped across a typical spread: it runs from about 0.01 directly underneath the B target to about 0.36 underneath the $SmB_6$ target. The desired composition of $SmB_6$ ($x = 0.14$) is obtained in the middle of the spread wafer. Fig. 1b shows the X-ray diffraction pattern (integrated in the chi direction) from the part of the spread film whose composition is $SmB_6$. A broad ring was seen in the area detector (see Supplementary I), which indicates the highly polycrystalline nature of the film. The pattern indicates that the film is preferentially oriented in the (001) direction, but peaks from other orientations such as (011) and (021) are also present. Upon scanning XRD across the entire range ($0.01 \leq x \leq 0.34$) we found that the predominant phase in the composition spread is always $SmB_6$, regardless of x (See Supplementary II). This is surprising because there are reports of several other intermetallic phases[29, 30] such as $SmB_4$, $Sm_2B_5$, etc.

To address this, we performed quantum mechanical parameterization of the Sm-B system to understand the phase formation and stability within the AFLOWLIB framework[31, 32] using the standard computational parameters as specified in Ref. 33. Fig. 2 shows, in blue line, the minimum formation free energy ($E_f$) at zero temperature. Based on the calculations, the low temperature thermodynamics would indeed dictate that the first precipitating phase should be $SmB_4$, having the minimum formation energy $E_f$ in the entire composition range, since $E_f$ represents the strength of the system to oppose phase separation. However, due to the hyper-thermal plasma process of sputtering, the Sm-B feedstock solution in the vapor phase starts off with an extremely large amount of intrinsic entropy (from translational degrees of freedom of vapor), and as a result, the entropic temperature $T_s$ of a compound (a measure of the entropy to overcome the formation energy) emerges as the key factor which determines the strength of the system to compensate for the entropy of the feedstock[33, 34]. Given the locations of phases in Figure 2, one sees that, from the point of view of the vapor, the most "accommodating" phase to nucleate is actually $SmB_6$ with its highest relative $T_s$ compared to the other phases. This follows from the fact that during nucleation, the entropy maximization will limit the flow of latent heat and therefore chooses the phase most capable of accepting and transforming entropy (from translational to configurational). Thus, the non-equilibrium nature of the co-sputtering process has served as the key enabler which allowed us to achieve the desired $SmB_6$ phase in the present synthesis method.

To further confirm the existence of $SmB_6$, we performed cross sectional transmission electron microscopy (TEM) and Raman spectroscopy on films with x near 0.14. Fig. 1c shows the high resolution cross sectional TEM images of films on a MgO substrate. Nanocrystalline $SmB_6$ grains with different orientations can be clearly seen. Fig. 1c2 and Fig. 1c3 show the enlarged images (after filtering) of two selected grains with (011) and (111) orientations, respectively. The lattice constant determined from the image is 4.14 Å, which is consistent with that of the reported bulk value. We also performed Atomic Force Microscopy (AFM) on our films and the typical rms roughness is less than 1.0nm for a 20μm by 20μm area.

Raman spectrum of a $SmB_6$ film on Si is taken at room temperature using a 532 nm wavelength laser. The result is shown in Fig 1d. Three modes $T_{2g}$, $E_g$ and $A_{1g}$ from $SmB_6$ (cubic structure with the Pm3m symmetry) are seen in addition to the Si substrate peak at 510 $cm^{-1}$ [Ref. 35]. The wavenumbers of these three peaks are consistent with those from the single crystals. They are first order Raman modes which involve the displacement of the boron octahedral. The Sm ion is at a site of inversion symmetry and cannot contribute to first order phonon Raman scattering. The fact that the widths of these three peaks are wider than the Si substrate peak also confirms that these films are polycrystalline.

We have carefully extracted different sections of the spread film with different compositions to carry out resistivity measurements. The sheet resistance $R_s$ (Fig. 3 left axis) and resistivity ρ (Fig. 3 right axis) vs temperature curves are plotted for three different boron concentration films with thickness about 100nm. We use sheet resistance here since at low temperatures the measured resistance arises from the surface state as discussed below. Comparing the three concentrations (x = 0.10, 0.14, and 0.22), the room temperature resistivity ρ(300K) increases with increasing boron concentration. At 300K, a typical resistivity for a stoichiometric $SmB_6$ film is around 300 μΩ cm, which is similar to that of bulk $SmB_6$ crystals[1, 2, 13, 16].

For single $SmB_6$ crystals, though the proposed topological property is yet to be verified, the bulk insulating and surface conductive nature have been established[8-11]. Our films show similar behaviors. As the temperature is decreased, all three films show increase in the resistivity with a particularly rapid rise below 50K. This behavior is consistent with the opening of the hybridization gap between the itinerant 5d band and localized Sm 4f band as found in bulk $SmB_6$ crystals[5-7]. Below 10 K, the sheet resistance displays a saturating behavior for the film with $x > 0.10$, but it continues to increase for the $x < 0.10$ sample. The saturation behavior, regarded as a mystery for decades, is now widely accepted as the signature of the emergence of the conducting surface state. We note that our saturated sheet resistance ($R_s$~50 Ω at 2 K) is very close to the reported sheet resistance of single crystals at low temperatures[9]. If the low temperature conduction is from the bulk of the film, one would expect a much larger sheet resistance because the films are three orders of magnitude thinner than typical crystals. Also if the conduction is from some trivial surface state, one would expect a very large sheet resistance due to grain boundary scattering because our films are nanocrystalline. The fact that we observe a very similar sheet resistance strongly supports that this low temperature conduction is from the surface and that this surface conduction is protected against conventional grain-boundary scatterings. This agrees with the predicted nontrivial nature of the surface states in $SmB_6$, in which the topologically protected surface states exist only at the interfaces between materials of different topology. Thus, the similar and robust surface conduction in our polycrystalline $SmB_6$ films compared to single crystals strongly supports the topological aspect of the surface states.

The resistivity ratio between 300 K and 2 K is about 1.7, which is substantially lower than that found in single crystals, where the ratio can be as large as $10^4$[Ref. 5-7]. This can be explained by a much larger surface to bulk conduction ratio for films. It has indeed been shown that when single crystals are thinned down from 200 μm to 37.5 μm, this ratio $\rho_{2K}/\rho_{300K}$ decreases from $10^4$ to less than 3000. A detailed study[15] has shown that the ratio has roughly a linear dependence on the thickness. Given the ratio value of 1.7, an extrapolation of this linear dependence gives a thickness of 242 nm, which is of the order of our film thickness (~100 nm). Of course, this linear relationship might not be valid down to 100 nm, but this rough agreement with the linear thickness dependence demonstrates that our thin films are already in a limit where the surface conduction at lowest temperature is comparable to the bulk conduction at room temperature. Therefore, we attribute the saturation and the small resistance increase in our films to the surface conducting state.

The Kondo nature of $SmB_6$ can be investigated by point contact spectroscopy (PCS) [11]. Fig. 4a (the bottom curve) shows a typical PCS curve of our $SmB_6$ films with a PtIr tip, exhibiting an asymmetric Fano line shape[36, 37]. When a superconducting tip is used for the contact (the top curve of Fig. 4a), inside the Fano dip and around the Fermi level a clear double peak feature appears, indicating the emergence of Andreev reflection, which takes place in a metal-superconductor junction. The fitting to the Fano line shape (blue curves) gives a Kondo gap of around 19 meV, which is similar to that observed in single crystals[11]. The red curve is the best fit to the Fano line shape and the Blonder-Tinkham-Klapwijk model[38], which gives $\Delta = 1.35$ meV, $Z = 0.57$ (barrier strength), $\Gamma = 1.0$ meV (broadening term). The gap value is consistent with that of Nb, and above its $T_c$, the spectrum only shows a Fano line shape with no Andreev enhancement. When the point contact force is varied, the barrier strength (Z) of the junction decreases with junction resistance, while clear signature of the superconducting gap remains (Fig. 4b). The co-occurrence of the Kondo feature and Andreev reflection in the $SmB_6$-superconductor junction clearly signifies the Kondo nature and the simultaneous existence of metallic states at the Fermi level in our films, which, although polycrystalline, reinforces the theoretical prediction that $SmB_6$ is a topological Kondo insulator. The observation of Andreev reflection implies that high transparency can be achieved in the superconductor/$SmB_6$ interface. Synthesis of $SmB_6$ thin films with true bulk insulation is an important step towards pursuing proximity induced topological superconductivity in $SmB_6$.

**Methods:**

**Fabrication.** $SmB_6$ and B targets were co-sputtered at dc 50 W and rf 100 W, respectively, in a high-vacuum combinatorial co-sputtering chamber. Boron is brittle and insulating, and its deposition rate is much lower than that of the $SmB_6$ target. The base pressure of the deposition chamber is typically 2 x$10^{-8}$ torr. The substrates used were MgO and 3-inch (001) oriented Si wafers with a 300 nm amorphous $SiO_2$ layer. Films on different substrates show consistent behavior. A physical shadow mask is placed over the wafer during the deposition to naturally separate the film into 2 mm x 2 mm squares. The typical argon pressure was 6.5 mTorr during the deposition, and the substrate was heated at 800 °C during the deposition and for 3 hours following the deposition in vacuum. The deposition thickness across the spread varied from 80 – 200 nm.

**Characterization and measurements** X-ray diffraction (XRD) of the spread is carried out with an area detector (Vantec500) using a Bruker D8 Discover with the CuKα line. Raman miscoscopy is done in a Horiba Jobin Yvon LabRam ARAMIS spectrometer using a 532 nm wavelength laser. Transport and point contact spectroscopy measurements are done in a Quantum Design Physical Property Measurement System (PPMS). Sheet resistance is measured in four point Van der Pauw configurations. Contacts are made by a wire bonder using Al wires. Sheet resistance is calculated by $R_s = V \pi / I / \ln 2$. Resistivity is calculated by $\rho = R_s t$ (thickness of the film)

Acknowledgement: This work was supported by ONR N00014-13-1-0635 and NSF DMR 1410665. IT and SC also acknowledge support by the Duke University Center for Materials Genomics.

Author contributions: J.Y., I. T. and R. G. conceived the project; J.Y. fabricated samples and did XRD, AFM and Raman studies; J. Y., X. Z, and J. S performed transport studies; L.L and X. P. performed TEM study; Y. J. carried out PCS studies; D. U. and S. C. performed the phase diagram calculations; J.Y., Y. J., D. U., S. C., I. T., R. G. wrote the manuscript. All authors contributed to discussions and gave comments to the manuscript.

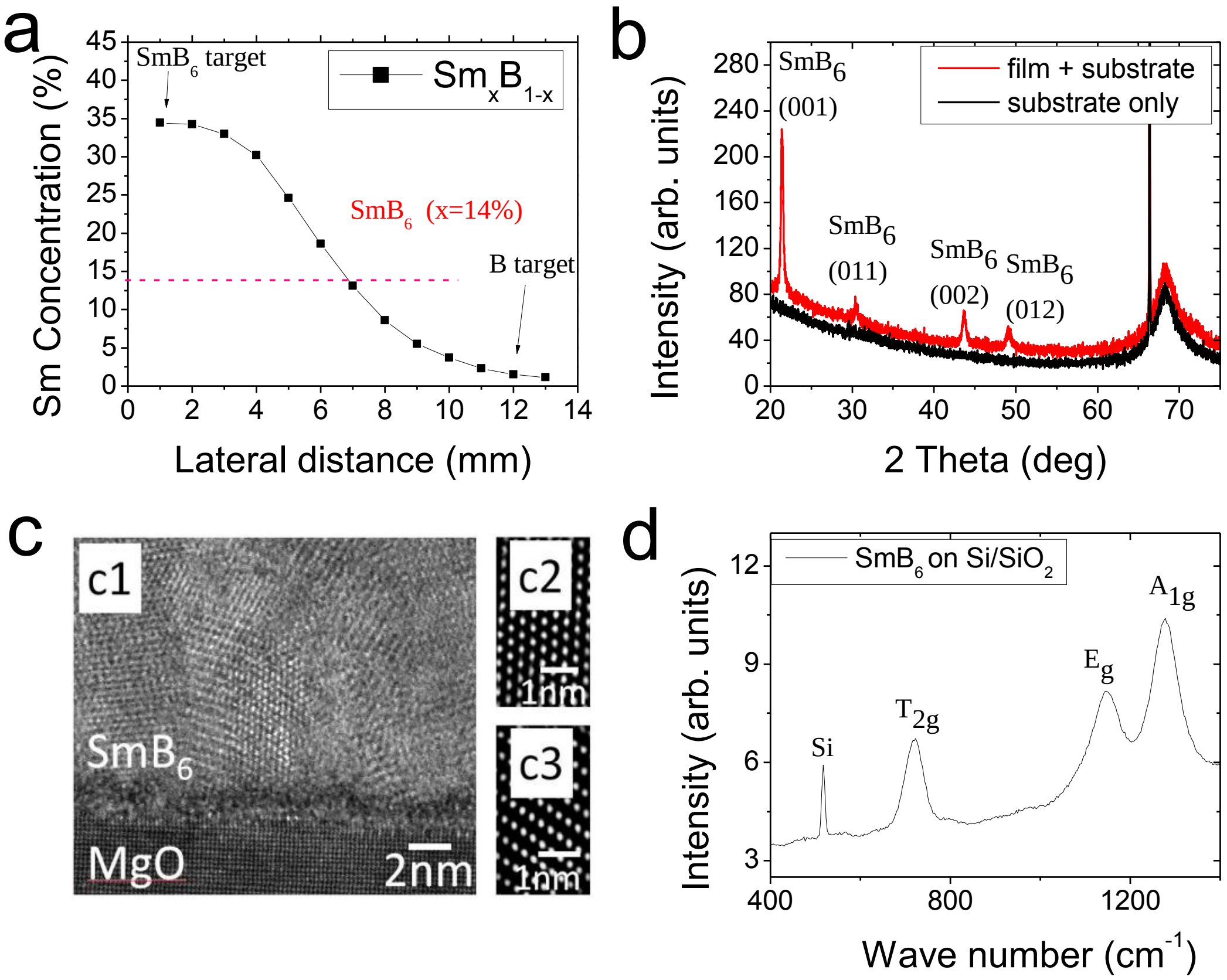

Figure 1: The fabrication and characterization of $SmB_6$. a, Samarium concentration x, measured by WDS, as function of lateral distance. b, Integrated X-ray diffraction intensity with aerial detector for the film with the substrate (in red) and with only the substrate (in black). The substrate is 300 nm $SiO_2$ on Si wafer. c, Cross-sectional TEM image on the interface between $SmB_6$ film and MgO substrate. (c1) raw image revealing textured structures with amorphous areas (c2) and (c3) are selected areas after filtering showing $SmB_6$ (011) and (111) planes. d, Raman spectrum of $SmB_6$ films on $Si/SiO_2$ substrate. The laser wavelength used is 532nm. Three modes $T_{2g}$, $E_g$, $A_{1g}$ from $SmB_6$ are clearly seen.

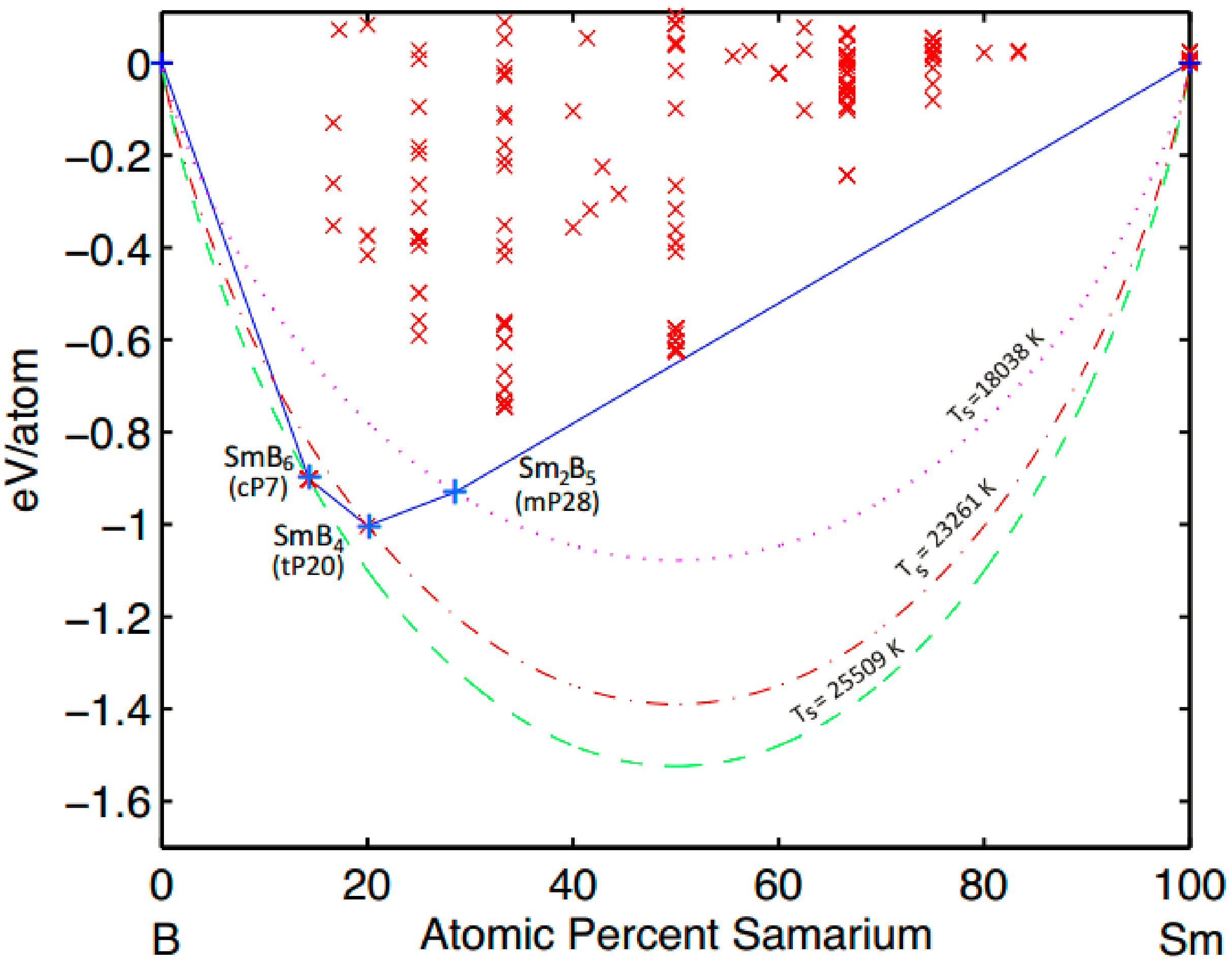

Figure 2: Free energy convex hull for B-Sm at zero temperature (solid blue line). Blue/red crosses indicate stable/unstable phases. Space groups are in parentheses below the stable compound stoichiometries. The three dashed curves indicate the entropic temperature envelopes[33, 34]. Cooling from the hyper-thermal plasma of sputtering it is actually $SmB_6$ with the highest $T_s$ which nucleates first.

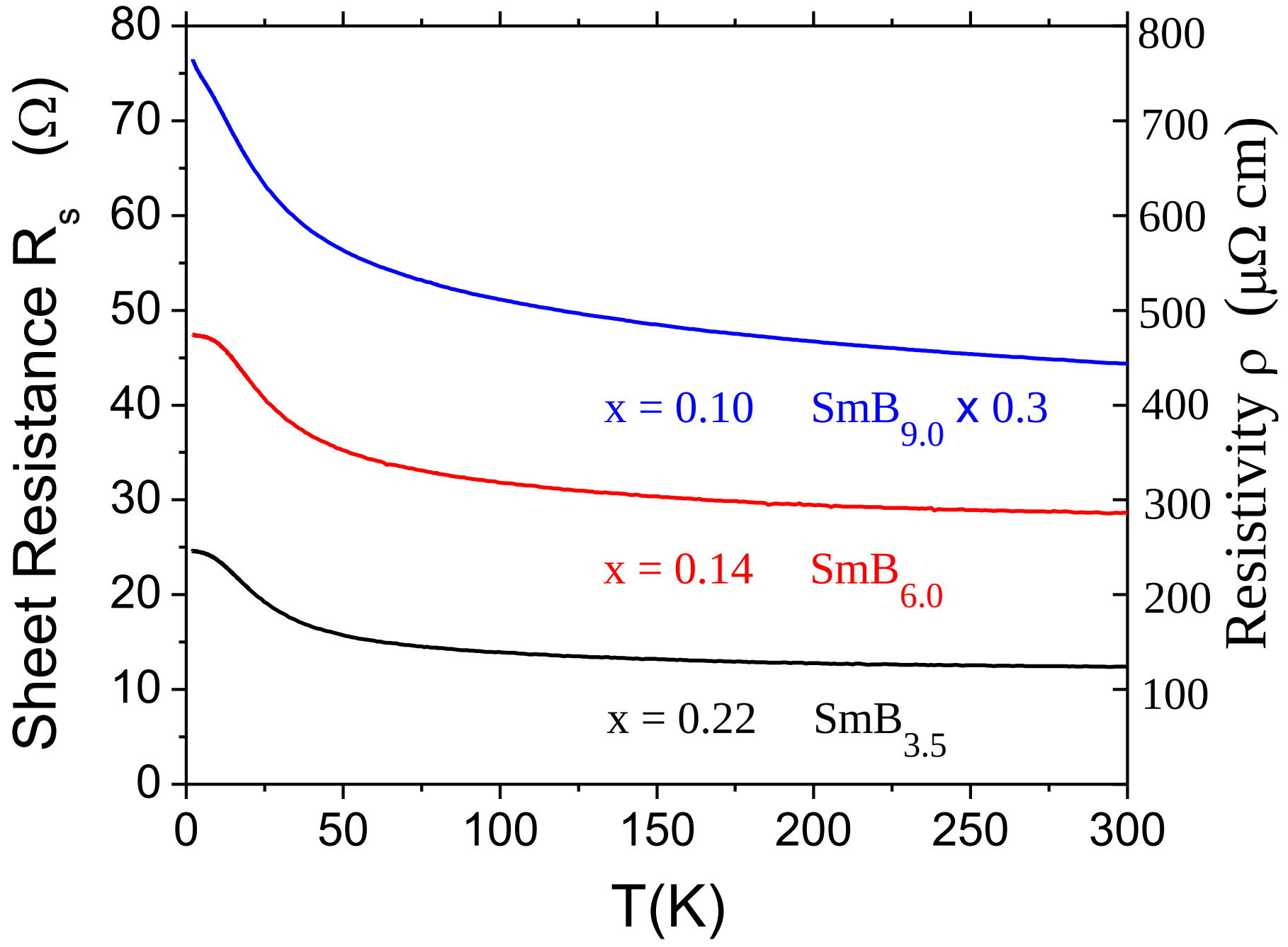

Figure 3: Temperature dependences of sheet resistances (left axis) and resistivities (right axis) for three $Sm_xB_{1-x}$ films with x=0.10, 0.16, 0.22 respectively. The films are about 100nm thick and resistances are measured using Van Der Pauw method (see Methods).

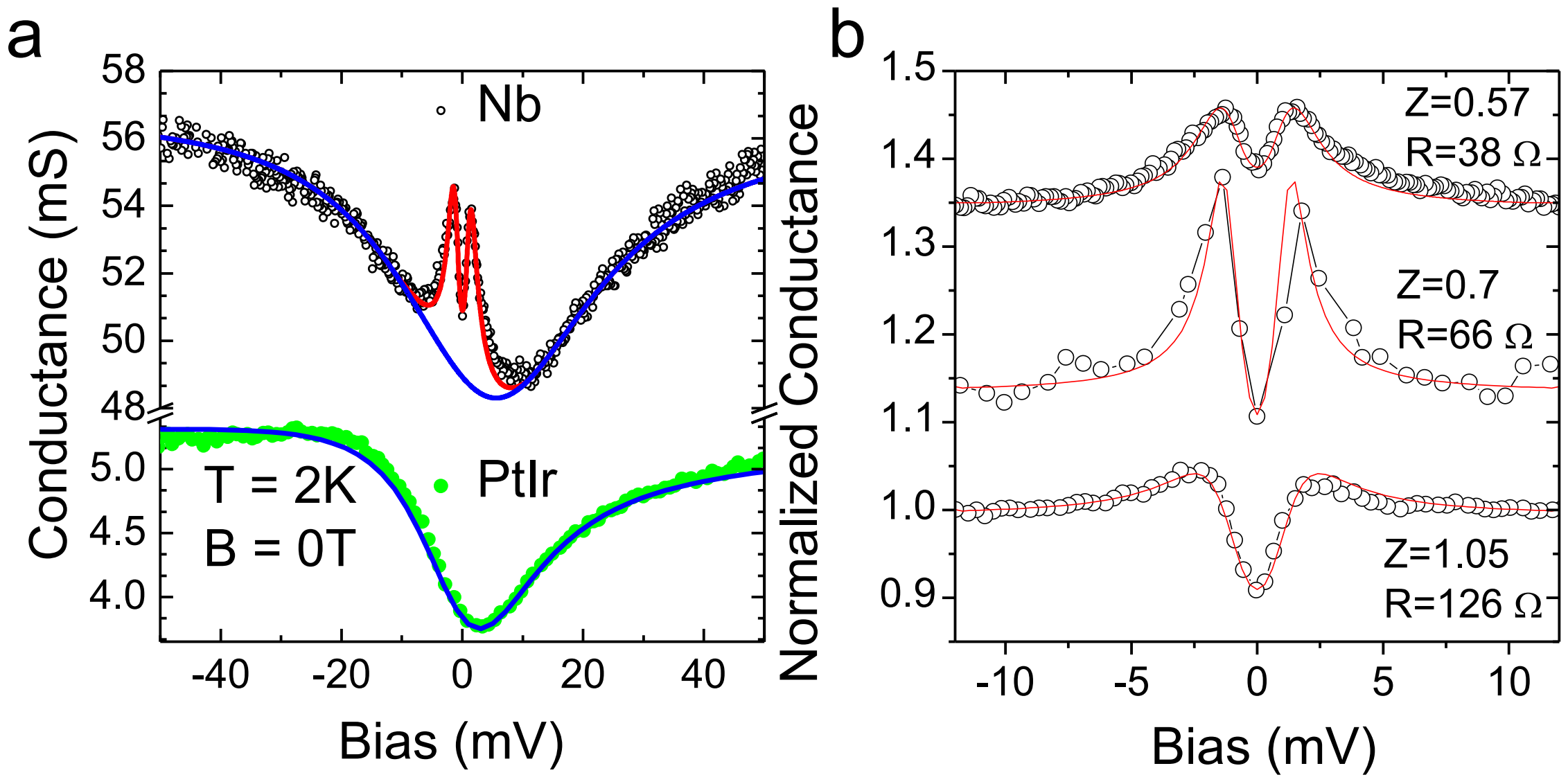

Figure 4: Point Contact Spectroscopy (PCS) on $SmB_6$ films. (a) PCS on the $SmB_6$ film with a PtIt tip (green filled squares) and a superconducting Nb tip (black open circles). The blue curves are the best fittings to the Fano line shape. The red curve is the fitting to both Fano line shape and modified BTK model. (b) Andreev reflection data (black open circles) of different Nb/$SmB_6$ junction resistances. Fano background is subtracted. The red curves are the fitting to the BTK model. The spectra here were taken at 2 K and in zero magnetic field. Details of BTK fit and data analysis are given in Supplementary III.

# Supplementary information

Jie Yong[1,2,*], Yeping Jiang[1,2,*], Demet Usanmaz[3], Stefano Curtarolo[3], Xiaohang Zhang[1,2], Linze Li[4], Xiaoqing Pan[4], Jongmoon Shin[5], Ichiro Tachuchi[1,5], Richard L. Greene[1,2]

[1]Center for Nanophysics & Advanced Materials, University of Maryland, College Park, Maryland 20742, USA

[2]Department of physics, University of Maryland, College Park, Maryland 20742, USA

[3]Department of Mechanical Engineering and Materials Science, Duke University, Durham, NC 27708

[4]Department of Materials Science and Engineering, University of Michigan, Anne Arbor, Michigan 48109, USA

[5]Department of Materials Science & Engineering, University of Maryland, College Park, Maryland 20742, USA

## I. 2D-XRD using an area detector

Polycrystalline films can be characterized using X-ray detector with an area detector. With a fixed angle between incident X-ray and the sample, the area detector can detect diffractions that come from the grains of different orientations in the sample. Polycrystalline films will show a broad ring (Fig. S1) in the detector while single crystal show a very broad spot (the incident angle must satisfy Bragg's law) which need to be avoided. Integration of the data along chi direction can be carried out to obtain θ - 2θ XRD plots (Fig.1b and Fig. S2)

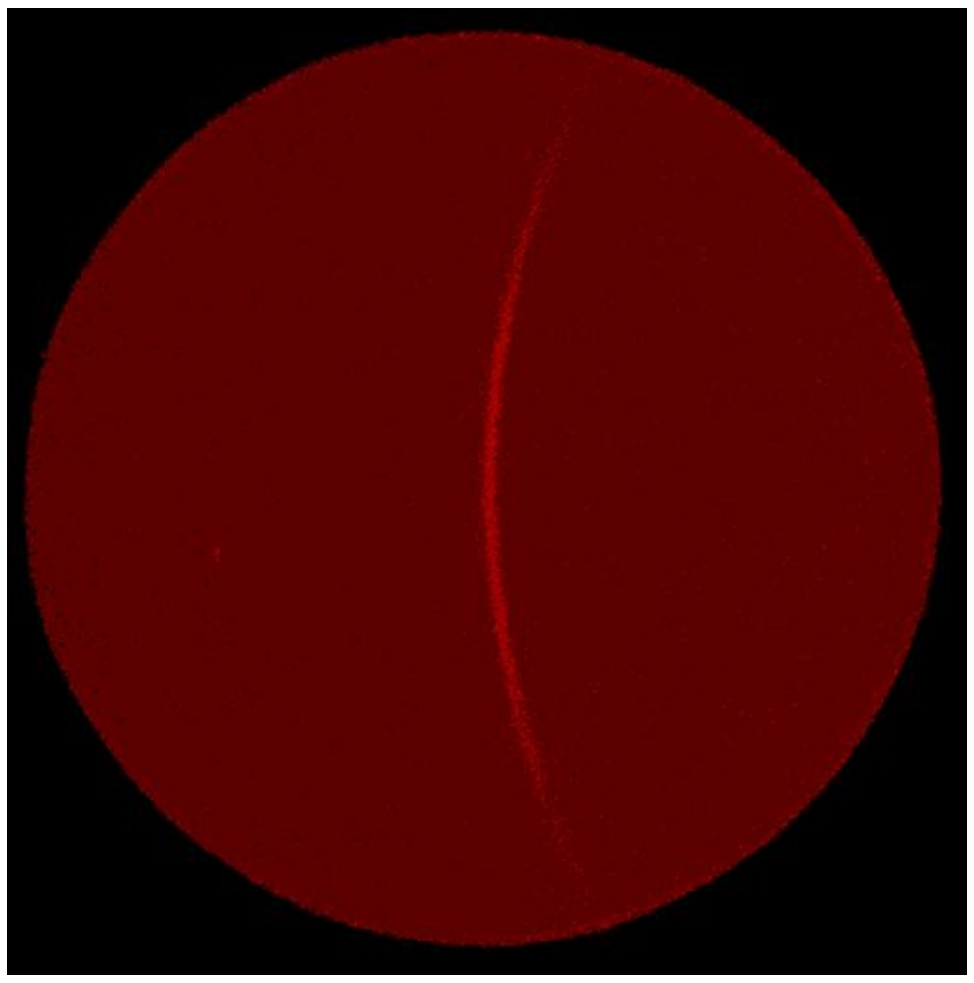

Figure S1: raw image of XRD area detector, the broad ring is from polycrystalline $SmB_6$ (001) peak.

# II. XRD spectra of $Sm_xB_{1-x}$ as a function of x

Figure S1 shows the spectra of $Sm_xB_{1-x}$ for seven different x. Only peaks from $SmB_6$ phase can be identified for any given x. This indicates $SmB_6$ is always the dominant phase in our growth setup, regardless of x. The main text explains why this is the case. Figure S3(a) shows the a small shift in $SmB_6$ (001) peak as x changes. This corresponds to a lattice constant change from 4.16 Å for small x to 4.12 Å for large x. This is because the higher the boron concentration (smaller x), the more boron atoms will fit into the lattice of much larger samarium atoms and the lattice constant increases. It is also interesting to note that the lattice change mostly occurs where samarium concentration is around 20% to 25%.

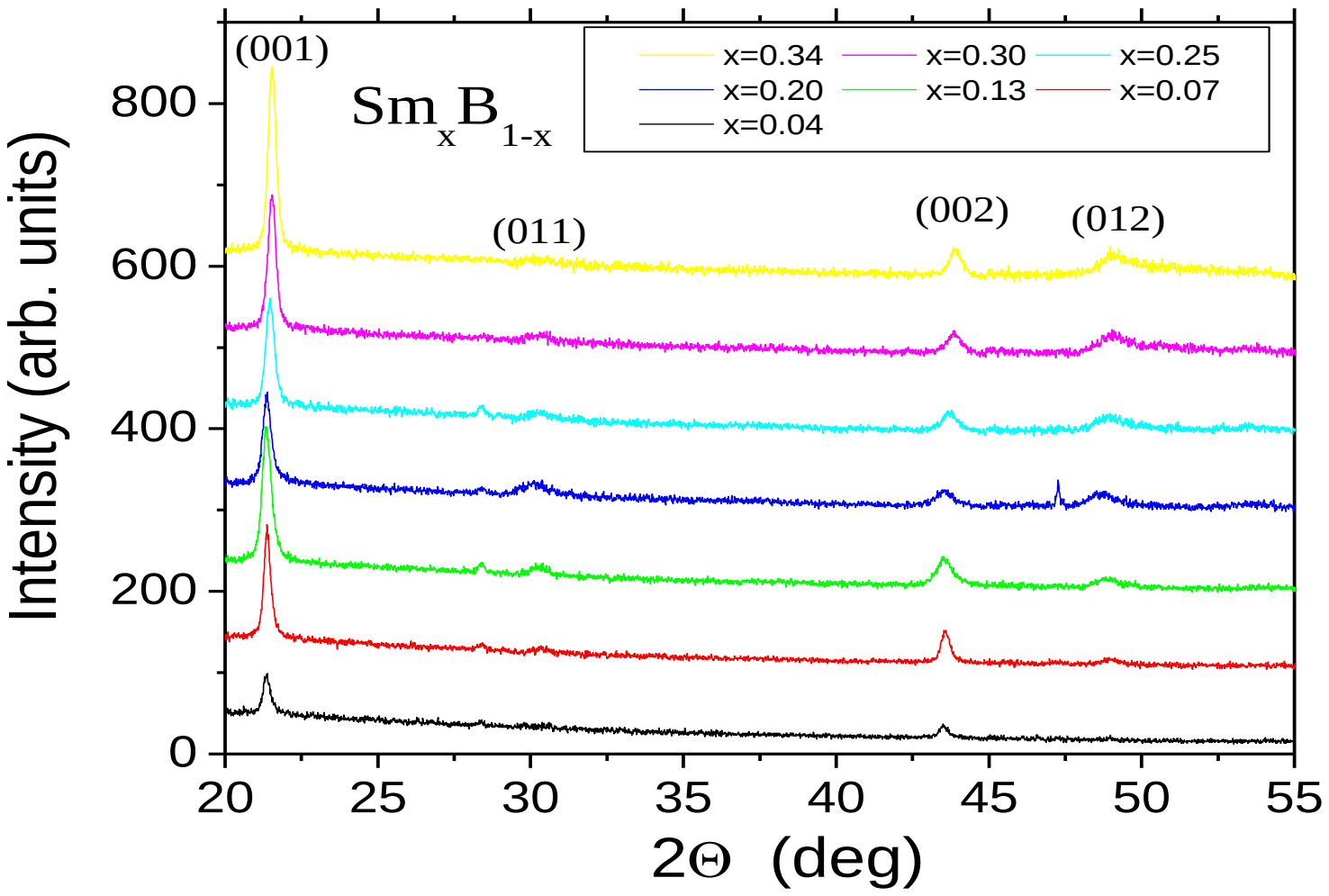

Figure S2: XRD spectra of $Sm_xB_{1-x}$ with different composition x. Vertical axis are shifted for clarity purposes.

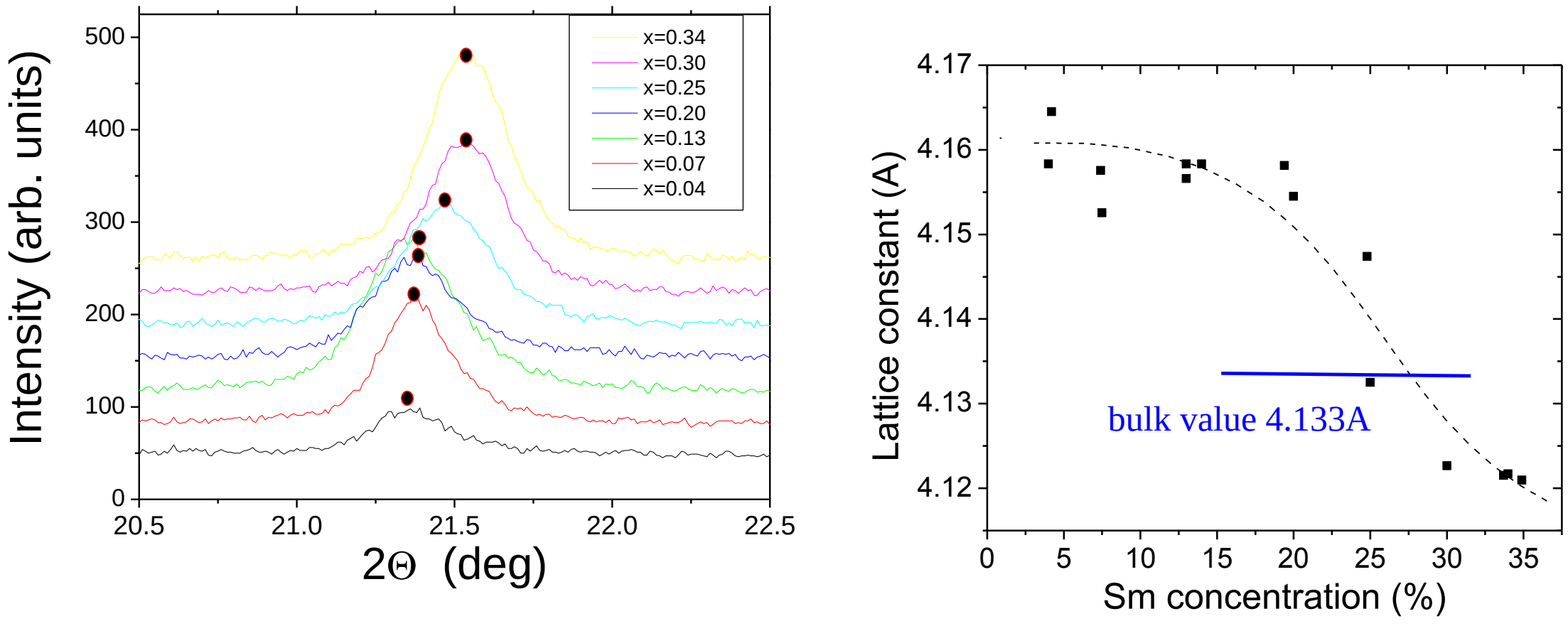

Figure S3: (a) 2-D XRD spectra on $Sm_xB_{1-x}$ films with different x showing $SmB_6$ (001) peak (purple dots) shift. Vertical axis is shifted for clarity purposes. (b) $SmB_6$ lattice constant, from fit of 2D XRD data, as a function of Samarium concentration in films. The dashed curve is a guide to the eye.

# III. Blonder-Tinkham-Klapwijk (BTK) fit

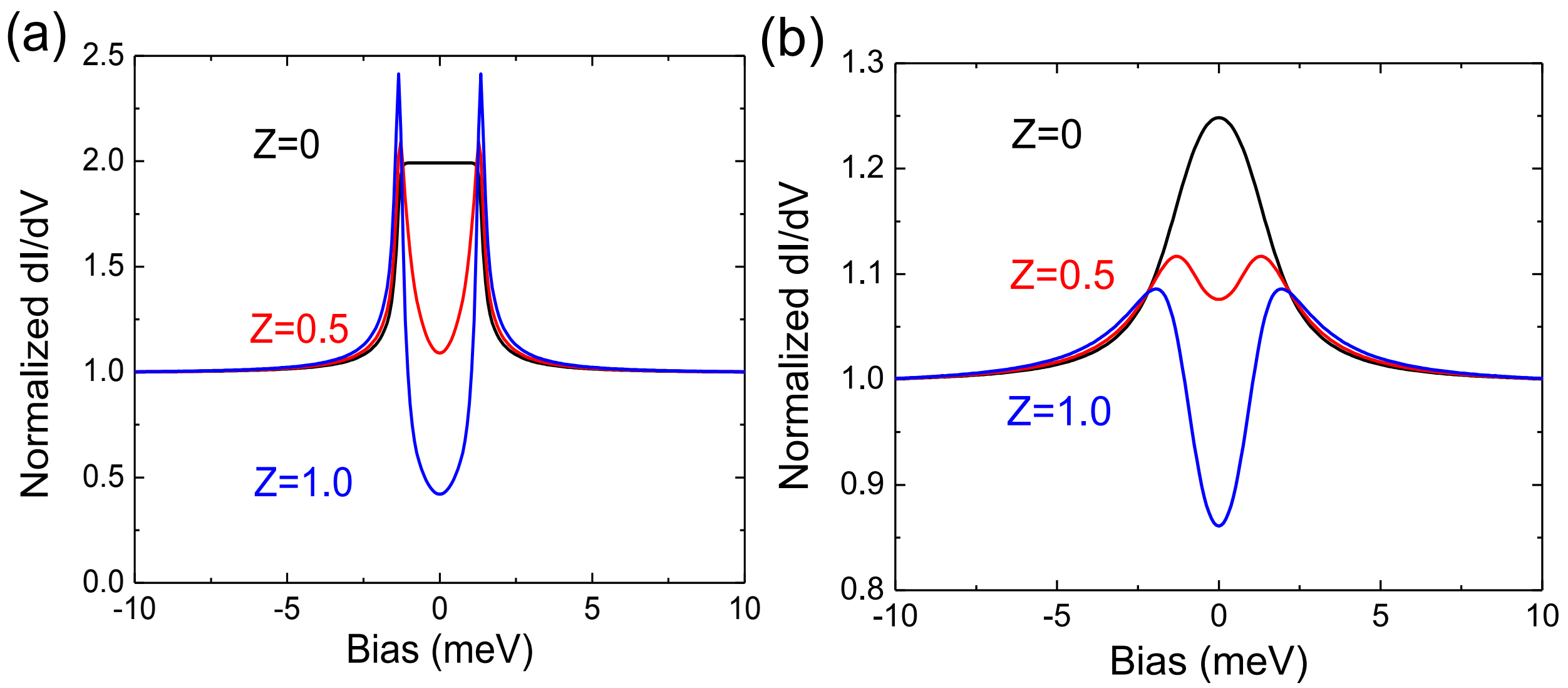

Figure S5： Calculated conductance curves using the modified BTK model. (a) Z = 0, 0.5, 1.0 with T, Δ and Γ fixed at 2.0 K, 1.35 meV and 0 meV. (b) Z = 0, 0.5, 1.0 with T, Δ and Γ fixed at 2.0 K, 1.35 meV and 1.0 meV.

The conductance $G_{NS}$ ($dI/dV$) as a function of the bias voltage $V$ of a N/S junction can be described by the BTK theory [ref. S1]:

$$G_{NS}(V) \propto \int_{-\infty}^{+\infty} \frac{\partial}{\partial V} [f(E-eV)][1+A(E)-B(E)]dE ,$$

where $A(E)$ and $B(E)$ are the probability of Andreev reflection and normal reflection, $f(E)$ is the Fermi distribution function. $A(E)$ and $B(E)$ can be calculated as follows,

$$\begin{aligned} A(E) &= a \cdot a^* \\ B(E) &= b \cdot b^* \\ a &= \tilde{u}\tilde{v}/\gamma \\ b &= -(\tilde{u}^2 - \tilde{v}^2)(Z^2 + iZ)/\gamma \\ \gamma &= \tilde{u}^2 + (\tilde{u}^2 - \tilde{v}^2)Z^2 \end{aligned} .$$

$Z$ represents the interfacial barrier strength, and the coherence factors $\tilde{u}$ and $\tilde{v}$ are

$$\tilde{u}^2 = \frac{1}{2}\left[1 + \frac{\sqrt{(E + i\Gamma)^2 - \Delta^2}}{E + i\Gamma}\right]$$
$$\tilde{v}^2 = \frac{1}{2}\left[1 - \frac{\sqrt{(E + i\Gamma)^2 - \Delta^2}}{E + i\Gamma}\right],$$

in which an imaginary component $\Gamma$ is introduced to describe a finite quasiparticle lifetime affected by inelastic scattering near the N/S interface. The inelastic scattering can result in a smearing effect on the conductance spectrum.

The effect of $Z$ and $\Gamma$ on the conductance spectrum is shown in Figure S5. For an ideal N/S interface with $Z = 0$ and $\Gamma = 0$, the spectrum has a flat top and the enhancement factor is 2 inside the superconducting gap where Andreev reflection takes place. As $Z$ increases, the conductance in the gap is suppressed. With a relatively large $\Gamma$, the enhancement at low Z is reduced and the double-peak structure in the case of a finite $Z$ is smeared.